\documentclass{phostproc}

\usepackage{url,hyperref,lineno,microtype,subcaption}
\usepackage{upgreek}
\usepackage{booktabs,caption,fixltx2e}	
\usepackage{aas_macros}
\usepackage[onehalfspacing]{setspace}
\usepackage{graphicx}

\title{Super-Nyquist Asteroseismology with Future Space Missions} 
\author{Hiromoto Shibahashi,$^{1}$ 
        Simon J. Murphy$^{2,3}$}

\affiliation{$^{1}$ Department of Astronomy, School of Science, The University of Tokyo, Tokyo 113-0033, Japan \\
$^{2}$ Sydney Institute for Astronomy, School of Physics, The University
of Sydney, NSW 2006, Australia\\
$^{3}$ Stellar Astrophysics Centre, Department of Physics and Astronomy, Aarhus University, 8000 Aarhus C, Denmark}

\shorttitle{Super-Nyquist Asteroseismology}
\shortauthors{H. Shibahashi \& S. J. Murphy}

\abs{We propose a photometric technique for future space missions that overcomes the problem of Nyquist aliases. These aliases result from typically long cadences of observation imposed by telemetry constraints. The proposed method is to introduce a periodic modulation to the sampling rate. Suitable combinations of the frequency and the amplitude of this modulation allow the true peaks to be distinguished from the aliases. We provide an analytical proof of the validity of this method and some demonstrations with simulated data.
We also propose to divide a long cadence into two unequal parts, aiming at reproducing the intrinsic amplitude spectrum of stars without a severe smearing effect due to long exposures.
The two exposures can be summed to recover the photon statistics if the user is interested in doing so. 
Based on these proposals, a specific recommendation for the PLATO mission is made to maximise its capability of photometry for asteroseismology, without serious interference with its other scientific missions.}

\begin{document}

\maketitle

\section{Introduction}
\label{Sec:1}

Photometric observations from space missions have provided us with uninterrupted data of ultra-high precision, taken over long timespans. These have enabled seismic observations of the deep layers of the Sun and the stars.

Due to limits on telemetry, however, the sampling rate (cadence) of space-based photometry is not as high as one would typically wish. Typical cases are the $\sim$30-min sampling of the long-cadence (LC) mode of the Kepler Space Telescope \citep{kochetal2010} and the full-frame images (FFIs) of the Transiting Exoplanet Survey Satellite (TESS) \citep{rickeretal2015}, and the 10-min sampling of the forthcoming PLAnetary Transits and Oscillations of stars (PLATO) mission. 
Many stars, including solar-like oscillators at the base of the red giant branch, some main-sequence A stars oscillating in pressure modes, and hot subdwarfs oscillating in gravity modes, have periods near to or shorter than this sampling rate, and consequently the Nyquist aliases of these oscillations complicate the data analysis and asteroseismic inference. 

{\it Kepler} observed a fixed star field continuously over four years whilst the spacecraft orbited the Sun. The cadence on-board the spacecraft was a fixed 29.45\,min, but the time stamps had to be corrected to the Solar System barycentre due to the spacecraft's orbit. By good fortune, the barycentric correction enabled the identification of true frequencies even higher than the Nyquist frequency \citep{murphyetal2012}. However, such long observational time spans might not be available in a general case, or orbital modulation of the sampling rate may not exist such as for the TESS continuous viewing zone \citep{murphy2015}.

In this paper, we first briefly describe how super-Nyquist asteroseismology was successfully carried out with the original {\it Kepler} mission, and then propose a new technique for future missions to overcome the problem of the Nyquist aliases.
We demonstrate its efficiency and show this technique to be useful for future space photometry missions.

\section{Discrete Fourier transform}
\label{Sec:2}

\subsection{General description}
\label{sec:2.1}

It is well known that the Fourier transform of discretely sampled data is expressed as a convolution of the Fourier transform of the original continuous function (in this case the stellar brightness) and the sampling window spectrum $W_N$, which is defined by
\begin{equation}
	W_N(\nu) := {{1}\over{N}} \sum_{n=1}^N \exp(-2\pi {\rm i}\nu t_n),
\label{eq:01}
\end{equation}
where $t_n$ ($n=1,2, ..., N$) denotes the sampling times. 
That is, the Fourier transform of the sample series consists of a superposition of the shape of the window spectrum, shifted by the frequencies involved in the original continuous function that describes the behaviour of the star 
(see Appendix\,1 for more detail).

\subsection{The case of a uniform cadence}
\label{sec:2.2}

\begin{figure} 
\centering
\includegraphics[width=0.97\columnwidth]{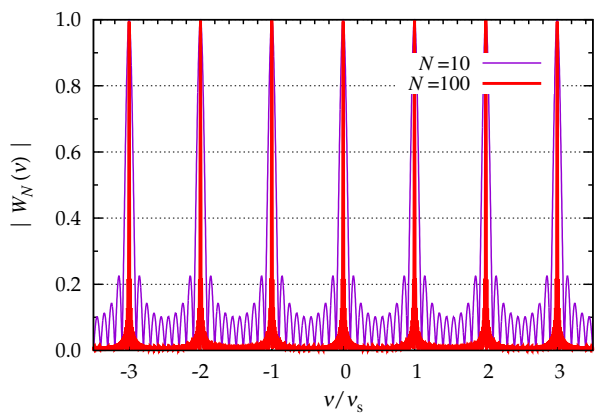}
\caption{The window spectrum in the case of sampling with a constant time interval.
With the increase of $N$, the window spectrum tends to be a shah function with peaks at $\nu/\nu_{\rm s}=j$, where $j=0, \pm 1, \pm 2, ...$ .}
\label{fig:01}
\end{figure}

If the light curve is sampled at a uniform cadence with a time interval $\Delta t$, then 
\begin{equation}
	t_n = t_0 + n\Delta t,
\label{eq:02}
\end{equation}
where $n = 1, 2, ..., N$,  and the window spectrum is reduced to 
\begin{equation}
	|\,W_N(\nu)\,| 
	=
	{{1}\over{N}}
	\left| {{\sin\{ N\uppi (\nu/\nu_{\rm s})\}}\over{\sin\{\uppi(\nu/\nu_{\rm s})\}}} \right| ,
\label{eq:03}
\end{equation}
where
\begin{equation}
	\nu_{\rm s} := {{1}\over{\Delta t}}
\label{eq:04}
\end{equation}
is the sampling frequency.
The window spectrum $|W_N(\nu)|$ has 
a series of sharp, high peaks at $\nu = j\,\nu_{\rm s}$, where $j=0, \pm 1, \pm 2, ...$ .
The peaks become sharper with the increase of $N$, and as $N\rightarrow\infty$
 the window spectrum approaches a shah function with peaks at $\nu=j\nu_{\rm s}$, where $j=0, \pm 1, \pm 2, \cdots$,
\begin{equation}
	\lim_{N\rightarrow\infty} |\,W_N(\nu)\,|  = {\sum_{j=-\infty}^{\infty}} \delta(\nu-j\nu_{\rm s}).
\label{eq:05}
\end{equation}

The convolution of the original continuous function with the window function thus causes each oscillation frequency to appear as a pair of peaks on each side of every $j\nu_{\rm s}$, whose separation in frequency from $j\nu_{\rm s}$ is equal to the true oscillation frequency, $\nu_{\rm true}$. That is, there occurs a peak at each frequency equal to $j\nu_{\rm s} \pm \nu_{\rm true}$.
These replicated peaks look identical, except for their frequencies, and we cannot uniquely determine which one is the true frequency peak.\footnote{Unless there is some additional external information. On the suggestion of \citet{chaplinetal2014}, \citet{yuetal2016} were able to determine the true oscillation frequencies in \textit{Kepler} targets at the base of the red giant branch by reference to the predicted frequency of maximum power, $\nu_{\rm max}$, from the asteroseismic scaling relations \citep{kjeldsen&bedding1995}.}

The peak in the range of $[0, \nu_{\rm s}/2]$ corresponds to the true frequency only if the true frequency\footnote{We take the true frequency to be positive.}
is lower than the Nyquist frequency $\nu_{\rm Nyq}$, defined as half of the sampling frequency.
Hence, when the plausible frequency range for the target pulsating star(s) is known in advance,
it is preferable to select the cadence so that this frequency range lies in the range $[0, \nu_{\rm Nyq}]$.  In the case of space missions, however, the cadence is determined before launch, largely due to telemetry constraints.

\subsection{Effect of barycentric correction}
\label{sec:2.3}

\begin{figure} 
\centering
\includegraphics[width=0.75\columnwidth]{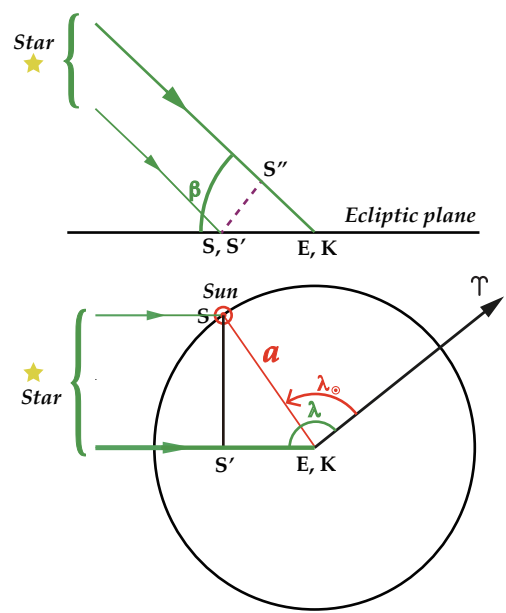}
\caption{
Geometric relations among {\it Kepler} (K), the Earth (E), the Sun (S) and the star. 
Top: schematic side view of the ecliptic plane, seen from a faraway point perpendicular to the direction to the star from {\it Kepler}.
Bottom: schematic top view of the ecliptic plane along the normal to that plane. 
We approximate {\it Kepler}'s Earth-trailing orbit as being circular with a radius of $a=1\,{\rm au}$ and in the ecliptic plane. Further, we approximate the angle between the direction to the star from {\it Kepler} as being close to the star's ecliptic longitude, $\lambda$. The geocentric ecliptic longitude of the Sun is $\lambda_\odot (t)$.
Let S$'$ be the intersection of the projection of the direction to the star on the plane and a line perpendicular to it through S. Then the length of $\overline{{\rm S}'{\rm K}}$ is $a\cos\{\lambda-\lambda_\odot(t)\}$, and the difference in the path length is $\overline{{\rm S}''{\rm K}} = \overline{{\rm S}'{\rm K}}\cos\beta = a\cos\beta\cos\{\lambda-\lambda_\odot(t)\}$, where $\beta$ is the ecliptic latitude of the star.
}
\label{fig:02}
\end{figure}

Barycentric corrections for {\it Kepler}'s orbit around the Sun modulate the formerly-regular time-stamps.
For the sake of simplicity, we ignore the difference between the barycentre and the heliocentre, and we approximate {\it Kepler}'s orbit as being circular and exactly in the ecliptic plane. 
Then the arrival time at the heliocentre is given by
\begin{equation}
	t_{\odot, n}=t_n - \tau\cos\left\{\lambda_\odot(t_n) - \lambda\right\},
\label{eq:06}
\end{equation}
where 
the second term represents the light time effect (see Fig.\,\ref{fig:02})\footnote{There is a typographical error in equation (14) of \cite{murphyetal2012}. The second term should be read with a minus sign; $t_{\odot,n}\equiv t_n-\tau\cos(\Omega t_n-\lambda)$.}.
Here, $\lambda_\odot (t_n)$ is the geocentric ecliptic longitude of the Sun, 
$\lambda$ is the ecliptic longitude of the target star, and $\tau$ is the amplitude of modulation, which equals the product of the orbital radius of the spacecraft measured in units of light travel time $(a/c)$ with the cosine of the ecliptic latitude of the star, i.e.,
\begin{equation}
	\tau = {{a}\over{c}}\cos\beta,
\label{eq:07}
\end{equation}
where $a$ and $c$ denote the orbital radius of {\it Kepler} and the speed of light, respectively, and $\beta$ is the ecliptic latitude of the target star.
The ecliptic longitude of the Sun is given, with the present assumptions, by
\begin{equation}
	\lambda_\odot (t) = {{2\uppi}\over{P_{\rm orb}}}(t_n - t_\gamma),
\label{eq:08}
\end{equation}
where $P_{\rm orb}$ is the orbital period of the spacecraft and $t_\gamma$ denotes the vernal equinox passage time of the Sun.

\begin{figure} 
\centering
\includegraphics[width=0.95\columnwidth]{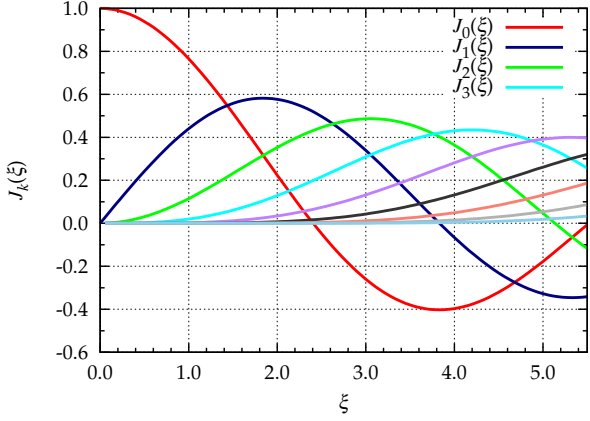}
\caption{The Bessel coefficients $J_k(\xi)$ with $k=0, ..., 8$ for $\xi=[0:5.5]$. }
\label{fig:03}
\end{figure}

Since the original {\it Kepler} mission continued observations of a fixed field for 4\,yr, then in equation\,(\ref{eq:06}) $t_N/P_{\rm orb} \simeq 4$ is sufficiently large that the window spectrum is reduced to  \citep{murphyetal2012} 
\begin{equation}
	|\,W_N(\nu)\,|
	=
	{{1}\over{N}}
	\sum_{k=-\infty}^\infty \Bigg|
	J_k(\xi) 
	{{ \sin [\, N\uppi\{(\nu+ k\nu_{\rm orb})/\nu_{\rm s}\} \,] }
	\over{ \sin [\,\uppi\{(\nu + k\nu_{\rm orb})/\nu_{\rm s}\}\,] }}
	\Bigg| ,
\label{eq:09}
\end{equation}
where $J_k(\xi)$ denotes the Bessel function of the first kind of integer order $k$, with 
\begin{equation}
	\xi = 2\uppi\nu\tau
\label{eq:10}
\end{equation}
and $\nu_{\rm orb}$ is the orbital frequency defined as
\begin{equation}
	\nu_{\rm orb} := {{1}\over{P_{\rm orb}}}.
\label{eq:11}
\end{equation}
Figure\,\ref{fig:02} shows the Bessel coefficients $J_k(\xi)$ with $k=0, ..., 8$ for $\xi=[0:5.5]$.

Note that $\nu_{\rm s}$ is much larger than $\nu_{\rm orb}$
and that $J_k(0)=0$ for $k\neq 0$.
The latter can be seen in Fig.\,\ref{fig:02}, where at $\xi = 0$ the central component ($J_0$) has an amplitude of 1 and all other components have zero amplitude.
Equation (\ref{eq:09}) means that the window spectrum $|W_N(\nu)|$ of the original {\it Kepler} mission consists of a singlet sharp peak at $\nu=0$ and multiplets with an equal spacing of $\nu_{\rm orb}$ around $\nu=j\nu_{\rm s}$, where $j=\pm 1, \pm 2, ...\,$. 
The width of any peak in the Fourier transform 
is $\sim \nu_{\rm s}/N = 1/ T_{\rm span}$, 
which is much smaller than the separation of peaks in the multiplets ($=\nu_{\rm orb}$), hence
each peak is well resolved.
The amplitude of each peak in the multiplets depends on the value of $j$, and it is given by $|J_k(2j\uppi\nu_{\rm s}\,\tau)|$.   

\begin{figure} 
\centering
\includegraphics[width=0.71\columnwidth,angle=90]{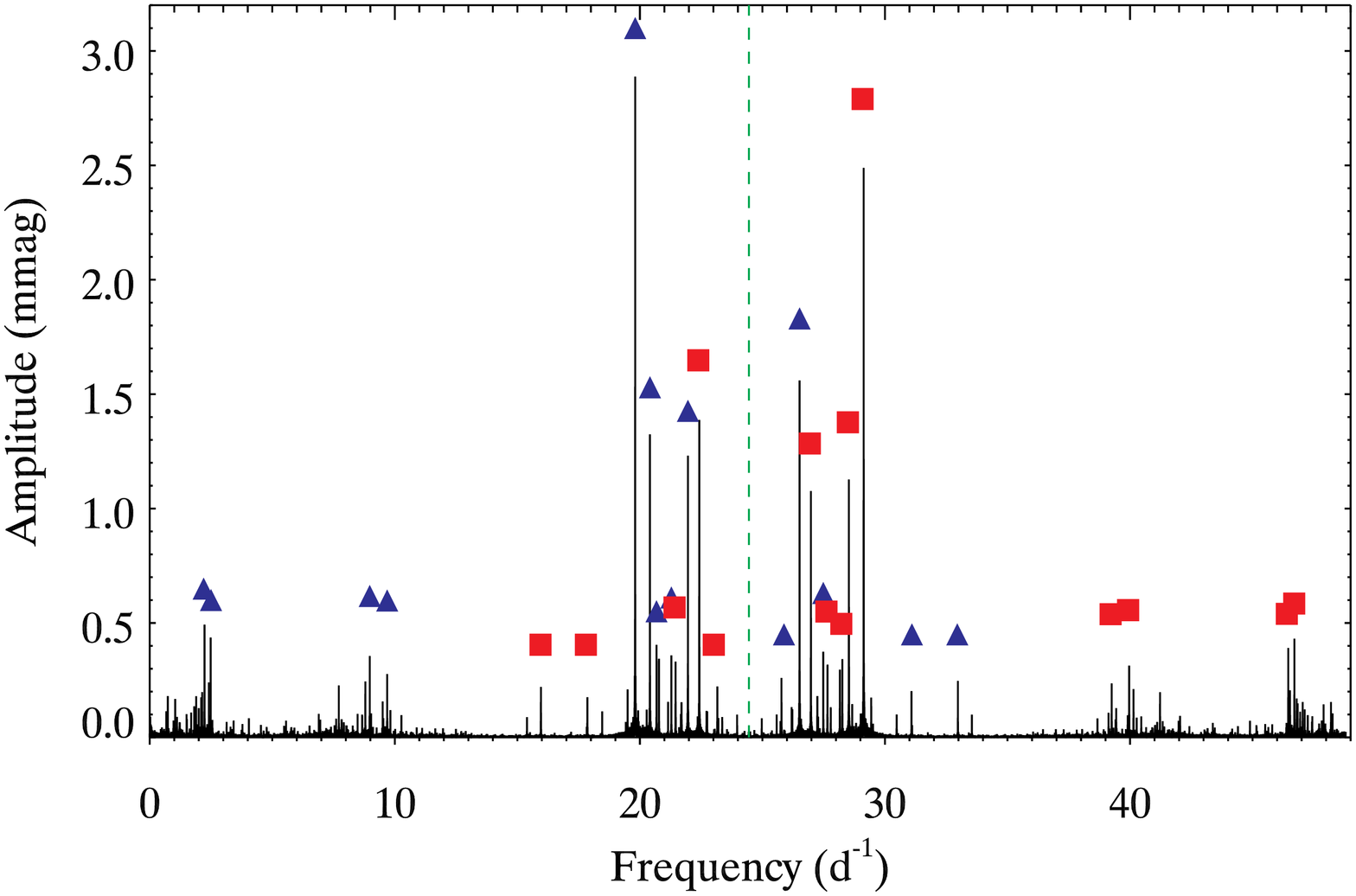}
\caption{
A Fourier transform of Q0--17, LC data for KIC\,4756040, calculated from 0\,d$^{-1}$ to the sampling frequency of the data at 48.9\,d$^{-1}$. The Nyquist frequency is indicated with a dashed green line. For some of the more prominent peaks, we indicate true peaks with blue triangles and aliases with red squares. This star has many true peaks above the LC Nyquist frequency as well as below it.
}
\label{fig:04}
\end{figure}

The Nyquist aliases are therefore split into frequency multiplets. For a true frequency $\nu_{\rm true}$ from the continuous function representing the stellar brightness, Nyquist aliases in the observed data will be found as multiplets, whose components are found at $j\nu_{\rm s} \pm \nu_{\rm true} \pm k\nu_{\rm orb}$ \citep{murphyetal2012}. 
The relative amplitudes of the components in each multiplet depend on the relation between the alias frequency and the sampling frequency, i.e. on the value of $j$, and are determined from the ratio of the Bessel coefficients $J_k(2\uppi j \nu_{\rm s} \tau)/J_0(2\uppi j \nu_{\rm s} \tau)$.

It follows that, for a true frequency at $\nu_{\rm true}$ where $j=0$, a single peak is observed. 
In the case of the long cadence mode of the original {\it Kepler} mission, since $2\uppi\nu_{\rm s}\tau \simeq 0.67$, the first Nyquist aliases at $1\nu_{\rm s} \pm \nu_{\rm true}$ (i.e. $j=1$) are essentially triplets, since the amplitudes of multiplets are dominated by a strong central component $J_0$ and weaker sidelobes ($J_1$). The second Nyquist aliases resemble quintuplets, 
and so on. In this way, the true peaks are readily distinguished from the aliases.

We illustrate this concept in Fig.\,\ref{fig:04}, using Q0--17 data\footnote{{\it Kepler} data are divided into seasons called Quarters (denoted Q$n$), each of around 93\,d duration, which is one quarter of the satellite's heliocentric orbit.} for the $\delta$\,Sct star KIC\,4756040, whose true peaks are distributed either side of the LC Nyquist frequency. Every true peak has an alias peak on the other side of the Nyquist frequency, but the alias appears as a triplet having a lower amplitude. With the original {\it Kepler} mission data, Nyquist aliases can be identified by amplitude even when the Fourier transform is quite dense with frequencies, as it is in this case.

\section{Proposal I : An artificial sampling modulation}
\label{Sec:3}

\begin{table*}
\caption{Characteristics of photometry from some of space missions.}
\centering
\begin{tabular}{lcccc}
\toprule
 & {\it Kepler} & K2 & TESS & PLATO \\
\midrule
\vspace{1.5mm}
Sampling period of LC & 30\,min & 30\,min & 30\,min & 600\,s \\
\vspace{1.5mm}
Observational time span & 4\,yr & 72\,d & 27\,d & 60\,d \\
\vspace{1.5mm}
Mission orbit & Earth trailing & Earth trailing & 2:1\, resonance with Moon & Earth--Sun L2 point\\
\vspace{1.5mm}
Light time effect & 190\,s & 500\,s & < 2\,s & $\leqslant$ 505\,s \\  
\vspace{1.5mm}
Reference & \cite{kochetal2010} & \cite{howelletal2014} & \cite{rickeretal2015} & http://sci.esa.int/plato \\
\bottomrule
\end{tabular}
\label{tab:01}
\end{table*}

Unlike the original {\it Kepler} mission, most space missions are not designed to observe a single field for as long as a few years (see Table\,\ref{tab:01}).
For example, the {\it K2} mission \citep{howelletal2014} had campaigns of 72-d duration, and TESS is observing fields for only 27\,d except for regions in or near the continuous viewing zones \citep{rickeretal2015}.
Then, $t_N/P_{\rm orb} < 1$ in most cases, hence we cannot expect that the multiplets in the window spectrum due to the barycentric correction for long cadence modes are resolved, so the Nyquist ambiguity remains.
What can we do then? 

\subsection{Mathematical support}
\label{sec:3.1}

We propose here to introduce an artificial periodic modulation in the architecture of the sampling:  
\begin{equation}
	t_{n}=t_0+n\Delta t + \tau_{\rm mod}\sin\left(2 \uppi n{{\nu_{\rm mod}}\over{\nu_{\rm s}}}\right), 
\label{eq:12}
\end{equation}
where $\tau_{\rm mod}$ is the modulation amplitude in units of time and $\nu_{\rm mod}$ is the modulation frequency. We propose that the exposure times themselves are fixed on board the spacecraft so that weights need not be applied to them afterwards.
We assume that the averaged sampling frequency $\nu_{\rm s}$ and the total timespan $T_{\rm span}$ are designed prior to actual observations.
The number of observations, $N$, is then also given as $N= T_{\rm span}\nu_{\rm s}$.
If $\Delta t = 30$\,min and the timespan is 30\,d, then $N:=T_{\rm span}/\Delta t =1440$.

\begin{figure} 
\centering
\includegraphics[width=1.0\columnwidth]{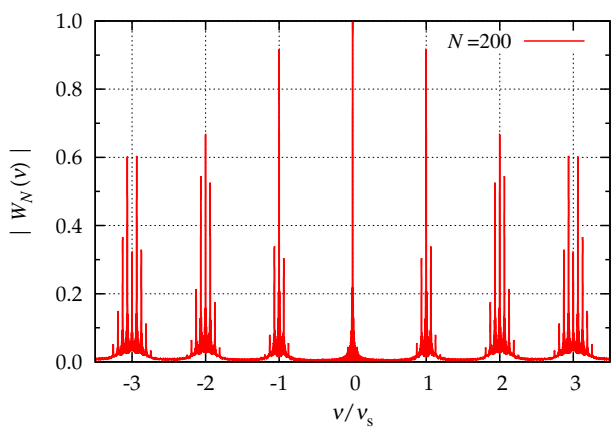}
\caption{Illustration of a window spectrum in the case of modulated sampling with $\xi=0.66$.
The peaks are distinguishable. 
}
\label{fig:05}
\end{figure}

\begin{figure*}[h]  
\centering
\includegraphics[width=0.8\columnwidth]{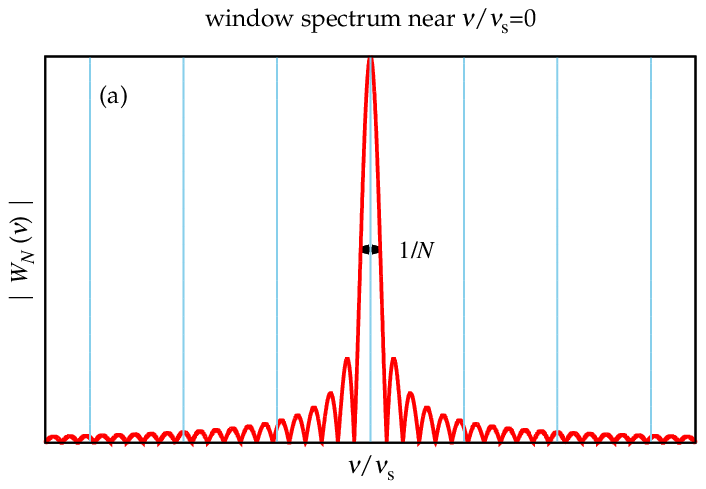}
\includegraphics[width=0.8\columnwidth]{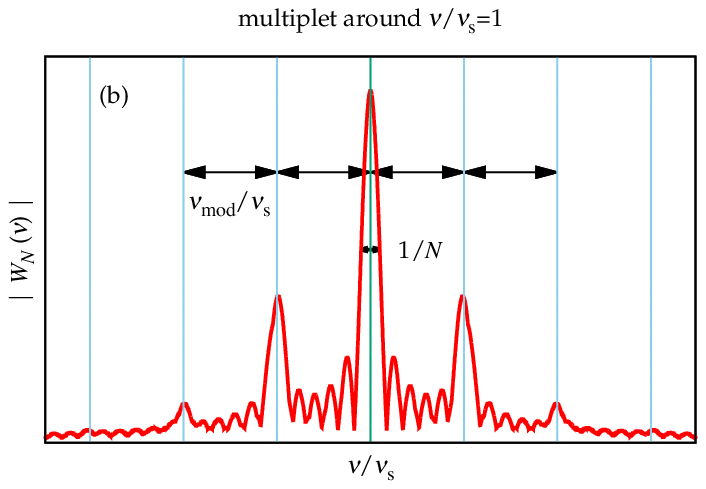}
\includegraphics[width=0.8\columnwidth]{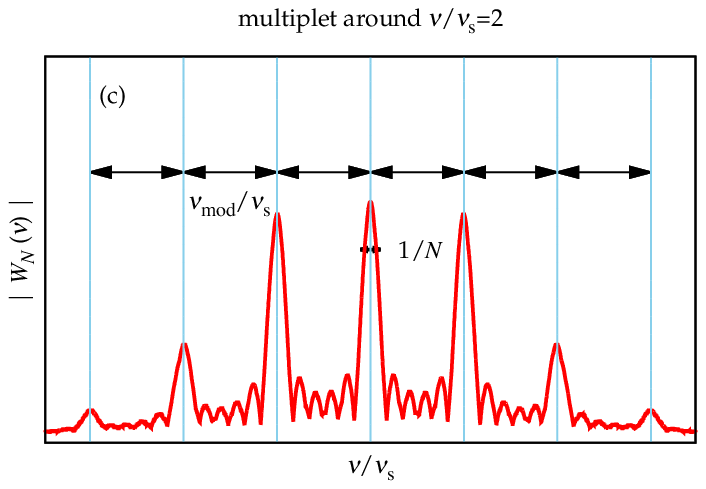}
\includegraphics[width=0.8\columnwidth]{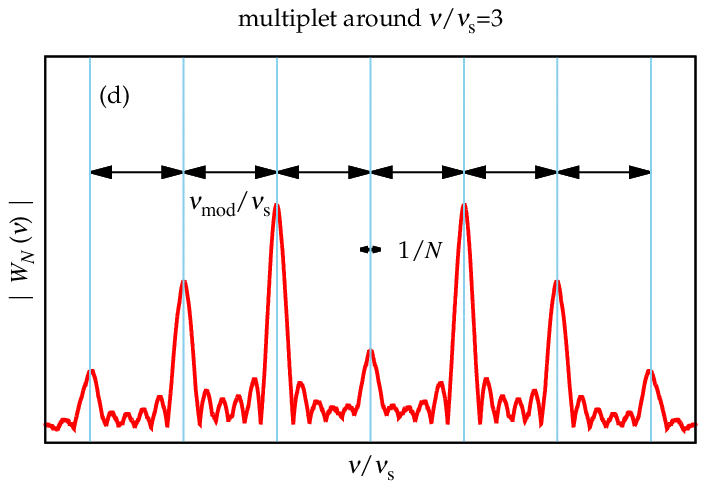}
\caption{Schematic illustrations for the window spectrum near {(a)} $\nu/\nu_{\rm s}=0$, {(b)} $1$, {(c)} $2$ and {(d)} $3$ plotted as functions of $\nu/\nu_{\rm s}$. In each panel, the central peak corresponds to the frequency $\widehat{\nu}=0$. The linewidth (full-width at half-maximum) of each peak is $1/N$, and the spacing in the multiplets is $\nu_{\rm mod}/\nu_{\rm s}$. 
The heights of peaks are dependent on $\nu_{\rm s}\tau_{\rm mod}$. All the plots shown here are the case of $\xi=0.66$.}
\label{fig:06}
\end{figure*}

The tuning parameters are the modulation frequency $\nu_{\rm mod}$ and the modulation amplitude $\tau_{\rm mod}$.
If we choose $\nu_{\rm mod}$ so that $\nu_{\rm mod} T_{\rm span}$ is sufficiently larger than unity,
we may have a window spectrum suitable to distinguish Nyquist aliases (see Appendix\,2 for more detail):
\begin{equation}
	|\,W_N(\nu)\,|
	=
	{{1}\over{N}}
	\sum_{k=-\infty}^\infty \Bigg|
	J_k(\xi) 
	{{ \sin [\,N\uppi \{(\nu+ k\nu_{\rm mod})/\nu_{\rm s}\}\,] }
	\over{ \sin [\,\uppi\{(\nu + k\nu_{\rm mod})/\nu_{\rm s}\} \,]}}
	\Bigg| ,
\label{eq:13}
\end{equation}
which is in the same form as the case of the original {\it Kepler} mission,
with 
\begin{equation}
	\xi = 2\uppi\nu\tau_{\rm mod}.
\label{eq:14}
\end{equation}

Then, the window spectrum consists of a single peak at $\nu=0$ and multiplets with an equal spacing $\nu_{\rm mod}$ around $\nu=  j\nu_{\rm s}$. 
With the increase of $N$, it tends to
\begin{equation}
	\lim_{N\rightarrow\infty} |W_N(\nu)|
	=
	{\sum_{j=-\infty}^{\infty}\sum_{k=-\infty}^{\infty}} J_k(j\xi_{\rm mod}) 
	\delta(\nu-j\nu_{\rm s}-k\nu_{\rm mod})
\label{eq:15}
\end{equation}
as far as $k \ll \nu_{\rm s}/ \nu_{\rm mod}$, 
where
\begin{equation}
	\xi_{\rm mod} := 2\uppi\nu_{\rm s}\tau_{\rm mod}.
\label{eq:16}
\end{equation}

Around $\nu=j\nu_{\rm s}$, the window spectrum is 
\begin{equation}
	|\,W_N(\widehat{\nu})\,|
	=
	{{1}\over{N}}
	\sum_{k=-\infty}^\infty
	\Bigg|
	J_k(j\xi_{\rm mod}) 
	{{ \sin [N\uppi \{\widehat{\nu}+ k({{\nu_{\rm mod}}\over{\nu_{\rm s}}})\}] }
	\over{ \sin [\uppi\{\widehat{\nu} + k({{\nu_{\rm mod}}\over{\nu_{\rm s}}})\} ]}}
	\Bigg| ,
\label{eq:17}
\end{equation}
where
\begin{equation}
	\widehat{\nu} := {{\nu - j\nu_{\rm s}}\over{\nu_{\rm s}}},
\label{eq:18}
\end{equation}
and in the limit of infinite data it tends to
\begin{equation}
	\lim_{N\rightarrow\infty} |\,W_N(\widehat{\nu})\,|
	=
	{\sum_{k=-\infty}^{\infty}} 
	\left|\,
	J_k(j\xi_{\rm mod}) 
	\,\right| 
	\delta \left( \widehat{\nu}-k\left({{\nu_{\rm mod}}\over{\nu_{\rm s}}}\right) \right)
\label{eq:19}
\end{equation}
as $N$ increases.
A schematic illustration is shown in Fig.\,\ref{fig:05} and zoom-in figures around $j=0, ... , 3$ are shown in Fig.\,\ref{fig:06}. 

\subsection{Choice of suitable parameters}
\label{sec:3.2}

The multiplet consists of the central peak at $\widehat{\nu}=0$ and the side peaks separated from it by $\nu_{\rm mod}/\nu_{\rm s}$. The width of the peaks is $\sim 1/N$. 
Hence, if we choose properly the modulation frequency for a given set of $\nu_{\rm s}$ and $N$ so that the separation is large enough compared to the peak width, each multiplet in the window spectrum is well resolved.
{With $\nu_{\rm mod}/\nu_{\rm s} \lesssim 1/N$, the full width at half-maximum of each peak in the multiplet is too large compared to the separation of peaks, and then the window pattern of each peak interfere with its neighbours. 
As already demonstrated by \cite{murphyetal2012} in their figure 6, a separation at least 
twice as wide as the Fourier peaks is needed, and a separation four times as wide has been shown to give adequate frequency resolution, both by \textit{Kepler} data and by the illustration in Fig.\,\ref{fig:06}. 
Then the minimum modulation frequency is 
	$\nu_{\rm mod} \gtrsim 2{{\nu_{\rm s}}/{N}}$, and 
\begin{equation}
	\nu_{\rm mod} \gtrsim 4{{\nu_{\rm s}}\over{N}} = {{4}\over{T_{\rm span}}}, 
\label{eq:20}
\end{equation}
is preferred,
i.e. the modulation period should be no less than a quarter of the time span of observations.}
In other words,
the window spectrum is fairly well approximated by equation (\ref{eq:15}) only if 
the above condition (\ref{eq:20}) is satisfied.

The height of the peaks in the $j$-th multiplet is dependent on the order of the side peaks $k$ and the combination of the modulation amplitude $\tau_{\rm mod}$ and the multiplet order $j$ through the argument of the Bessel coefficient, $j\xi_{\rm mod}$. 
The value of $\xi_{\rm mod}$ may be arbitrarily chosen, unless it is $\xi_{\rm mod} \sim 0$ leading to difficulty in detecting any side peaks in aliases. 
It is also inconvenient, if the central peak of the first alias multiplet is too low to determine its frequency accurately. Further, it is practically inconvenient if there appear too many side peaks in the first alias to identify easily the central frequency of the first alias multiplet. 
With the help of Fig.\,\ref{fig:02}, we are rather negative about choosing $\xi_{\rm mod} \gtrsim 2$.   
On the other hand, it is convenient to have the first alias and the second alias look like a triplet and a quintuplet, respectively, to identify these aliases morphologically and determine their central frequencies accurately.
Such a situation is realised if we tune $\tau_{\rm mod}$ so that $\xi_{\rm mod} \simeq 2/3$, that is, if 
	$\tau_{\rm mod} \simeq {{1}/{(3\uppi\nu_{\rm s})}}$,
or more roughly,
\begin{equation}
	\tau_{\rm mod} \simeq {{1}\over{10\nu_{\rm s}}}.
\label{eq:21}
\end{equation}
So, our recommendation is $\tau_{\rm mod} = \Delta t/10$. 

It was very fortunate that the barycentric correction for the original {\it Kepler} mission satisfied these suitable conditions of equations (\ref{eq:20}) and (\ref{eq:21}) by chance.

\subsection{Some remarks}
\label{sec:3.3}

The timespan $T_{\rm span}$ should be long enough to resolve the intrinsic dense frequency spectrum.  
To resolve the two adjacent modes separated from each other by $\delta\nu$, we need to set $T_{\rm span} > 1/\delta\nu$. In the case of $\delta\nu = 1\,\mu{\rm Hz}$, $T_{\rm span} > 12\,{\rm d}$ is required.
To resolve the rotational splitting of a star, $T_{\rm span}$ should be longer than the rotation period of the star. Furthermore, to catch a binary feature by pulsational frequency or phase modulation, $T_{\rm span}$ should be longer than the orbital period {\citep{shibahashi&kurtz2012, murphyetal2014}}.

{
In the case of solar-like oscillations, which are stochastically excited by turbulence in a convective envelope, each mode has intrinsically a Lorenzian profile in the power spectrum, whose full width at half maximum is determined by the damping rate of oscillations, $\mathit\Gamma$.
Hence the width of peaks in the Fourier transform is the larger of two options: either the damping rate or the reciprocal of the timespan. A typical width is of the order of one to a few $\mu{\rm Hz}$ \citep{chaplin&miglio2013}. 
If the width of peaks in the Fourier transform is governed by the damping rate rather than the timespan, we need to choose the modulation frequency so that 
\begin{equation}
	\nu_{\rm mod} \gtrsim {{4}{\mathit\Gamma}}.
\label{eq:22}
\end{equation}
}

We always need to make the barycentric correction. If $T_{\rm span}$ is as long as the orbital period of the spacecraft around the Sun, the barycentric correction becomes important. If the targets lie at appropriate ecliptic latitude such that $\tau = (a/c)\cos\beta$ is large enough to be useful, then the barycentric correction leads to suitable modulation 
and we no longer need any artificial modulation. Otherwise, the artificial modulation period determined by equation (\ref{eq:20}) is much shorter than the orbital period of the spacecraft, so that the spacings of these two modulations are easily distinguishable.

A gradual periodic modulation whose amplitude is not too large (i.e. $\tau_{\rm mod} \ll \Delta t$) enables a spacecraft to retain a high duty cycle. This is preferable over, say, intermittent cessation of observation in order to generate irregularly sampled data, which would be an (undesirable) alternative way to raise the effective Nyquist frequency \citep{eyer&bartholdi1999,koen2006}. In addition, because: (i) the spacecraft is automated; (ii) the duty cycle is to be kept as high as possible; and (iii) the exposure times as equal as possible,  there is no small and random delay between observations that might otherwise modulate the sampling frequency \citep{koen2010}.

A spacecraft that orbits the Sun (even if it does so by orbiting the Earth) does not necessarily have a large light-time effect, since this effect 
also depends on the ecliptic latitude of the targets. It is for this reason that TESS targets in the continuous viewing zone do not benefit from the prescription of super-Nyquist asteroseismology described here. However, a different sampling strategy was proposed to facilitate super-Nyquist asteroseismology for TESS \citep{murphy2015}.

\section{Nyquist frequency is no longer the detection limit}
\label{Sec:4}

We have shown in the previous section that the Fourier transform of the window spectrum with a periodically-modulated sampling interval consists of multiplets around the averaged sampling frequency. It should be stressed here that the shape of the multiplets is determined by just two parameters: one is the ratio of the modulation frequency to the sampling frequency and the other is the product of the modulation amplitude and the sampling frequency. The Fourier transform of the discretely sampled data consists of multiplets, whose shape is identical to that seen in the window spectrum, around the frequencies of $j\nu_{\rm s}\pm \nu_{\rm true}$, where $\nu_{\rm true}$ denotes the true frequency of the target.  
Only the peak at the true frequency appears as a single peak, while the other Nyquist aliases are multiplets. Hence, by finding the unique singlet, we can distinguish the true pulsation frequency from its Nyquist aliases. Note also that this feature applies not only for the case of pulsations below the Nyquist frequency but also for the case of those higher than the Nyquist frequency.  The Nyquist frequency is no longer the upper limit of frequency determination in such cases, hence super-Nyquist asteroseismology is possible by tuning the two parameters. This has been mathematically proved here.

\section{Proposal II :  Dividing LC into two unequal exposures}
\label{Sec:5}

\begin{figure} 
\centering
\includegraphics[width=0.9\columnwidth]{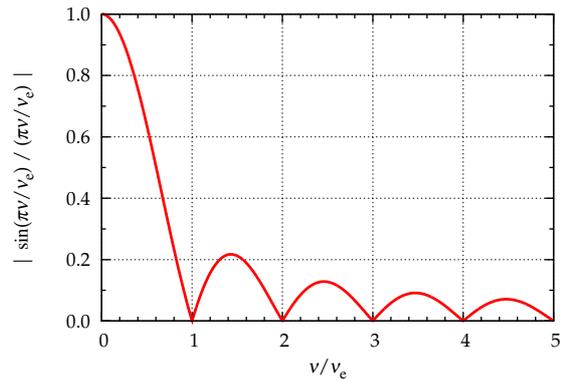}
\caption{
The attenuation factor due to the finiteness of the exposure time as a function of the ratio of the frequency to the reciprocal of the exposure time, $\nu_{\rm e}$. At frequencies $\nu=n\nu_{\rm e}$, the observed amplitudes become null. 
}
\label{fig:07}
\end{figure}

The actual photometric data are obtained by integrating the detected photons over a pre-set exposure time. Due to this integration, the observed amplitude is attenuated depending on the ratio of the oscillation frequency to the reciprocal of the exposure time, $\nu_{\rm e}$. 
The attenuation factor $\eta(\nu/\nu_{\rm e})$ is given by the cardinal sine function or sinc function \citep{chaplinetal2011,murphy2012a} 
\begin{equation}
	\eta (\nu/\nu_{\rm e}) = \left| {{\sin(\uppi \nu/\nu_{\rm e})}\over{\uppi\nu/\nu_{\rm e}}} \right|
\label{eq:23}
\end{equation}
as is shown in Fig.\,\ref{fig:07}.
If the exposure time constitutes the majority of the cadence of observations, as in the case of {\it Kepler}'s LC mode,
amplitude reduction is substantial except for very low frequency oscillations. 
In particular, the oscillations in the frequency range near to the sampling frequency (and its integer multiplets) are significantly attenuated, hence it is impossible to deduce the intrinsic amplitudes, which contain clues about the physics of excitation mechanisms.

To avoid this difficulty, we propose here to split the cadence into two exposure runs, a short exposure such as 1\,min and a long exposure with the remaining sampling time. The short exposure measurement with little attenuation is done every cadence, and is followed immediately by the long exposure. 
Since the number of photons detected during one short exposure is much smaller than in the long cadence, the photon noise is much higher. However, this noise level  is not more than in the short cadence, and the reduction in amplitude smear compensates for the higher noise.   

Both time-series should be downloaded. By summing up later the photon fluxes measured by the two exposure runs, we recover the photon flux which would be measured in the original long cadence. 
This allows for simpler analysis of transits and long-period oscillations without the complication of applying weights to the time series. The only drawback is the increased demand on telemetry, but the benefits justify this.

\section{Numerical simulation}
\label{Sec:6}

\begin{figure}
\centering
\includegraphics[width=0.9\columnwidth]{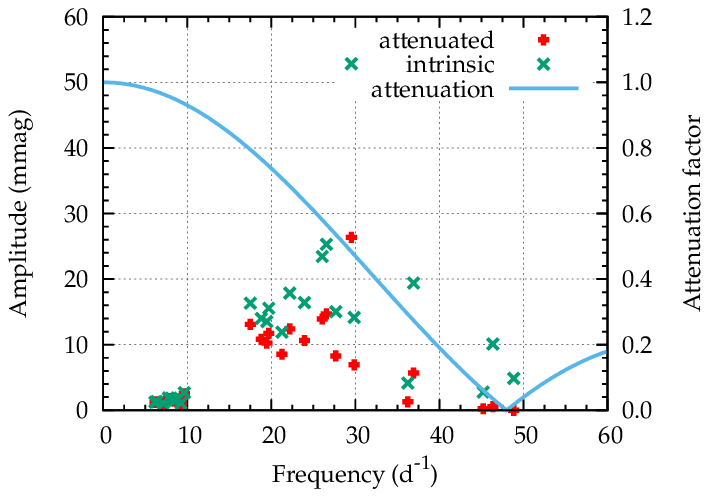}
\includegraphics[width=0.9\columnwidth]{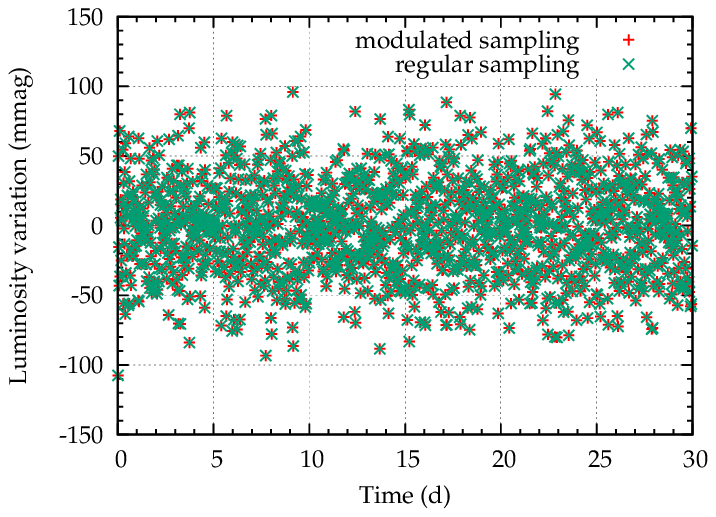}
\caption{Top: the amplitudes and frequencies of the simulation data. The intrinsic amplitudes are shown with the green crosses. Due to 29.45-min exposure, the observed amplitudes are substantially reduced as shown with red asterisks. Bottom: simulated luminosity as time series data. The luminosity variation without sampling modulation is also plotted to demonstrate how inconspicuous the modulation is. 
}
\label{fig:08}
\end{figure}

We demonstrate the proposals made in the previous sections by carrying out numerical simulations with artificial data constructed from 
a set of eigenfrequencies of pulsation modes with spherical degree $0$\,--\,$3$ representing a realistic model of a $1.9\,M_\odot$ main-sequence star. We set their amplitudes so that they represent pulsations seen in a $\delta$\,Sct variable. 
White noise is assumed.
We assume that $\Delta t = 29.45\,{\rm min}$ 
and $T_{\rm span} = 30\,{\rm d}$. Thus, $N=1440$.  The tuning parameters adopted in this simulation are (i) $\nu_{\rm mod} = 0.125\,{\rm d}^{-1}$, corresponding to a modulation period of $\sim T_{\rm span}/4$, and (ii) $\tau_{\rm mod} = 190\,{\rm s}$. Thus, $\xi_{\rm mod}=0.66$.

\subsection{Long exposure case}
\label{sec:6.1}

We prepare two sets of artificial data of time series photometry. In the first set,
we set the exposure time to be 100\% of the 29.45-min sampling interval, 
similar to that of \textit{Kepler} LC data. 
We assumed the noise level is 0.2\,mmag per point. 
This long exposure induces amplitude reduction, represented by the sinc function with a period of 29.45\,minutes shown in the top panel of Fig.\,\ref{fig:08}, and the attenuated amplitude distribution is shown by red crosses in the bottom panel of Fig.\,\ref{fig:08}.
The time series thus generated is shown in the bottom panel of Fig.\,\ref{fig:08}, and compared to the case with no sampling modulation to show that the modulation is inconspicuous.

\begin{figure*}
\centering
\includegraphics[width=0.8\columnwidth]{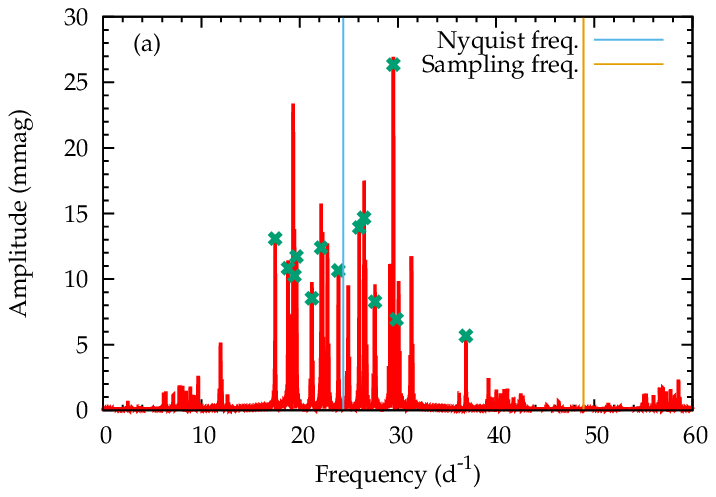}
\includegraphics[width=0.8\columnwidth]{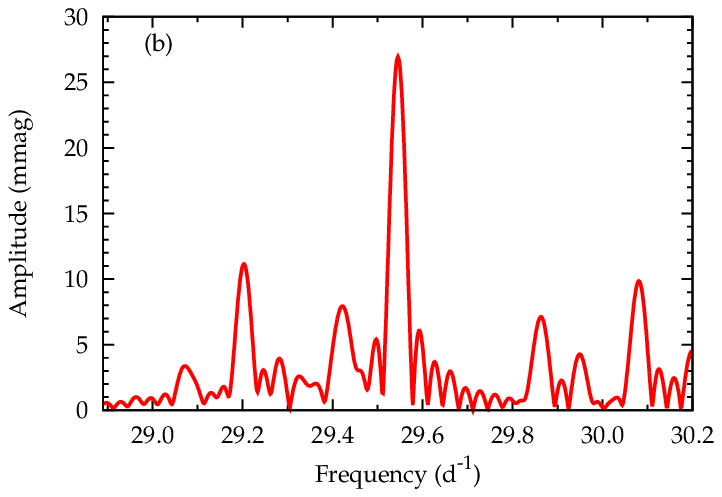}
\includegraphics[width=0.8\columnwidth]{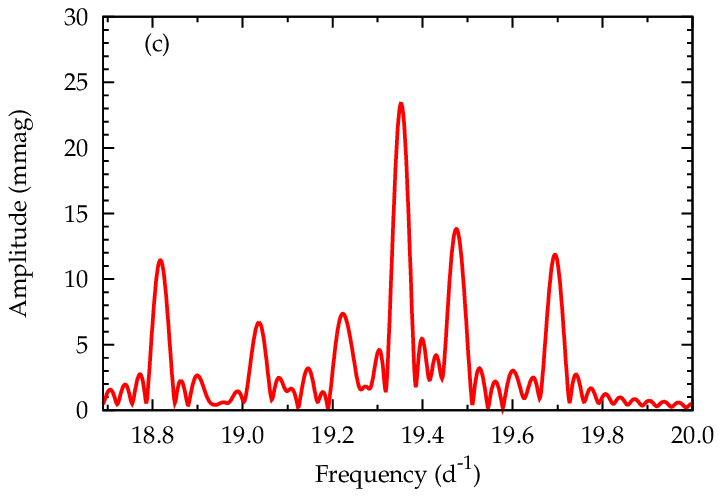}
\includegraphics[width=0.8\columnwidth]{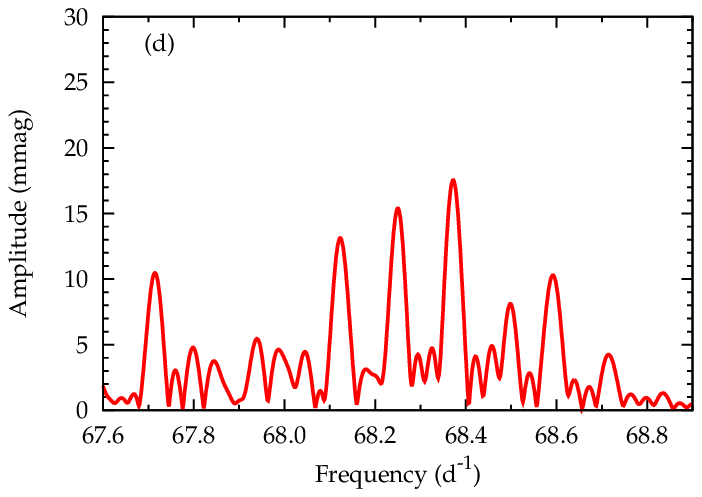}
\caption{{(a)} The amplitude diagram obtained with the Fourier transform of the simulation data with 30-min exposure in the frequency range of $0-60\,{\rm d}^{-1}$. 
{(b)} A zoom-in plot near the frequency range of the highest peak. 
{(c)} A zoom-in plot around the frequency range of the first order alias of the highest peak. 
{(d)} A zoom-in plot around the frequency range of the second order alias of the highest peak.}
\label{fig:09}
\end{figure*}

The amplitude spectrum obtained by the Fourier transform of the time series data is shown in Fig.\,\ref{fig:09}a. The amplitudes are attenuated near the sampling frequency (48.9\,d$^{-1}$). The amplitudes of the input data are marked with the green crosses for some high amplitude modes. Several modes have higher frequencies than the Nyquist frequency (24.45\,d$^{-1}$), and the amplitudes of some of their aliases appearing below the Nyquist frequency are larger even than many of the other true modes. So, if one performs the Fourier analysis only in the frequency range lower than the Nyquist frequency, those aliases would be misidentified as true modes. 

Figure\,\ref{fig:09}b shows a zoom-in plot around the frequency range of the highest peak. The highest peak is a singlet, so we can conclude that this is a true mode. 
Figure\,\ref{fig:09}c shows a zoom-in plot around the frequency range corresponding to the difference between the sampling frequency and the true one, as deduced from the frequency of the singlet in the top right panel. A triplet with a separation equal to the modulation frequency is found and hence we conclude that this peak is a first alias ($j=\pm1$) of the true mode. 
On the other hand, {panel (d)} illustrates a quintuplet, hence a second order alias ($j=\pm2$).

\begin{figure}[h!] 
\centering
\includegraphics[width=0.8\columnwidth]{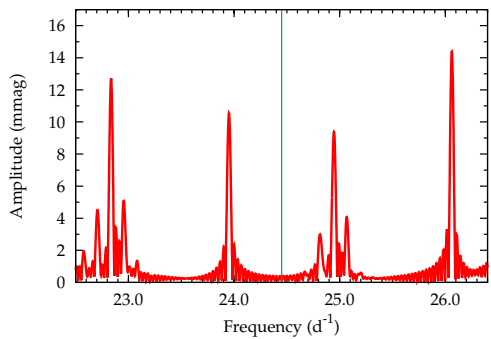}
\caption{The amplitude diagram obtained with the Fourier transform of the simulation data in the frequency range near the Nyquist frequency. Four peaks having the same order of amplitudes are found. They would be hard to judge true modes or aliases without introducing the modulated sampling. The Nyquist frequency is indicated with a thin green line.}
\label{fig:10}
\end{figure}

This simulation clearly demonstrates that we can identify correctly the order of aliases and distinguish the true frequencies from their aliases by setting suitable modulation parameters. 
 
Another demonstration is shown in Fig.\,\ref{fig:10}, in which the frequency range near the Nyquist frequency is zoomed in. There are four peaks, of which two are multiplets: one below the Nyquist frequency and the other above the Nyquist frequency. 
The triplet structure and the peak amplitudes are both consequences
of the modulated sampling cadence. Hence we can identify which ones are the true peaks. 

\begin{figure*} 
\centering
\includegraphics[width=0.8\columnwidth]{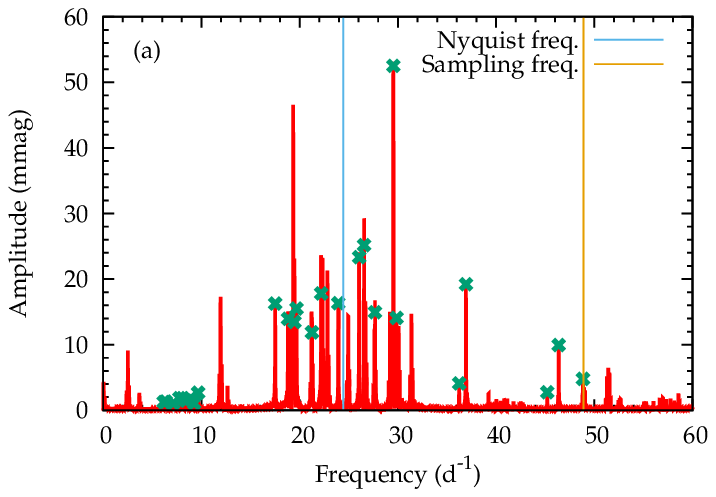}
\includegraphics[width=0.8\columnwidth]{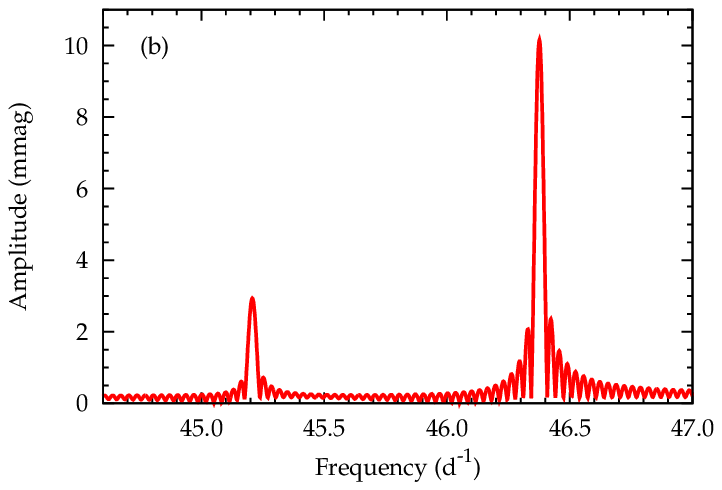}
\includegraphics[width=0.8\columnwidth]{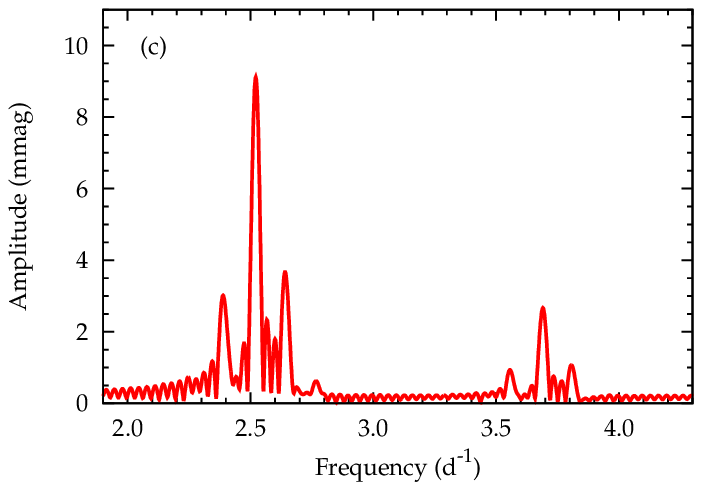}
\includegraphics[width=0.8\columnwidth]{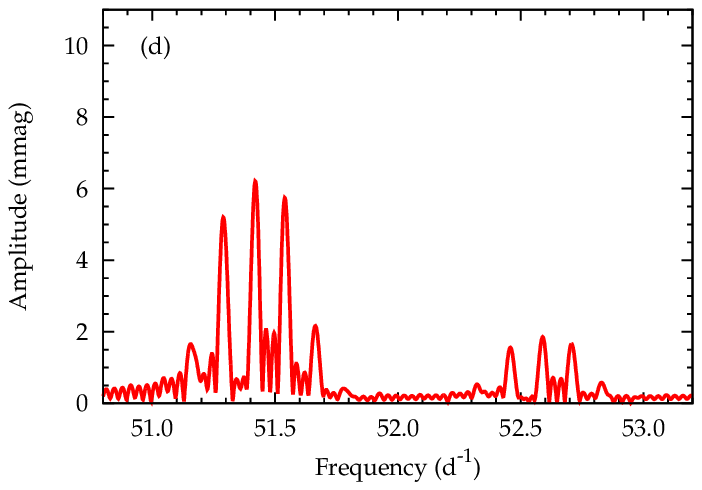}
\caption{
{(a)} The amplitude diagram obtained with the Fourier transform of the simulation data with 3-min exposure in the frequency range of $0$ -- $60\,{\rm d}^{-1}$. 
{(b)} A zoom-in plot near the frequency range of $45$ -- $47\,{\rm d}^{-1}$. 
{(c)} A zoom-in plot around the frequency range of $2$ -- $4\,{\rm d}^{-1}$, showing the first order aliases of the modes shown in {panel (b)}. 
{(d)} A zoom-in plot around the frequency range of the second order alias of the peaks shown in {panel (b)}.
}
\label{fig:11}
\end{figure*}

\subsection{Short exposure case}
\label{sec:6.2}

In the second set of simulations, aiming at short exposure runs with little attenuation, we assume that 
{the sampling time of 29.45-min
is divided into two runs, one with only 3-min short exposure and the other with 26.45-min long exposure, and these two series are downloaded separately so that the measurement with the whole sampling time is recovered later.}
The modulation parameters are the same as in the previous case. 

{Here, we analyse the short exposure runs.}
The number of photons detected during one short exposure time is only 1/10 of the previous case, so the noise is higher by a factor of $\sqrt{10}$. 
The Fourier spectrum of the generated time series is shown in Fig.\,\ref{fig:11}a. 
The intrinsic amplitudes of the modes are marked with the green crosses for some high amplitude modes, and they are well reproduced in the Fourier spectrum of the simulated data.
As seen in this diagram, the intrinsic spectrum of the star is well reproduced in the Fourier transform of observations obtained with a much shorter exposure compared to the averaged sampling time interval.
{Panel (b)} shows the Fourier spectrum around the frequency range of $45$\,--\,$47\,{\rm d}^{-1}$, much higher than the Nyquist frequency. Each of the two high peaks is a singlet, and then the both are identified as the true modes. 
On the other hand, the peaks in {panel (c)} around the frequency range of $2$\,--\,$4\,{\rm d}^{-1}$, which is the difference between the sampling frequency and the range of $45$\,--\,$47\,{\rm d}^{-1}$, are triplets, and then {they are identified not as a true mode but as a first order alias of the true peaks shown in the top right panel.} The second order aliases can also be seen as quintuplets just above the sampling frequency, as shown in {panel (d)}.   

This simulation with short exposure demonstrates that the true frequencies are correctly identified even if they are higher than the Nyquist frequency and that the intrinsic spectrum of the model can be then well reproduced by dividing the spectrum produced from the identified true peaks by the attenuation factor. The latter point is an advantage which is not realized with the previous case of $\nu_{\rm e}=\nu_{\rm s}$.

\section{Recommendation for PLATO}
\label{Sec:7}

PLATO is due to be launched in 2026 by the European Space Agency (ESA) to find transiting exoplanets and to conduct asteroseismology. A set of 24 cameras will continuously monitor stars fainter than $V=8$ with a cadence of 25\,s or 600\,s for 60\,d.
The longer cadence is shorter than the LC mode of Kepler, but short-period variables may still oscillate more rapidly than the sampling rate.

We propose here to introduce a periodic modulation to the sampling rate in order to overcome the Nyquist aliases and to maximise capability of PLATO.
From equations (\ref{eq:20}) and (\ref{eq:21}) with $T_{\rm span}=60$\,d and $\nu_{\rm s}=1/600\,{\rm s}^{-1}$, 
we adopt the parameter values of 
	$\nu_{\rm mod} = {{1}/{15}}\,{\rm d}^{-1}$
and 
	$\tau_{\rm mod} = 60\,{\rm s}$.

\begin{figure}[t!] 
\centering
\includegraphics[width=0.9\columnwidth]{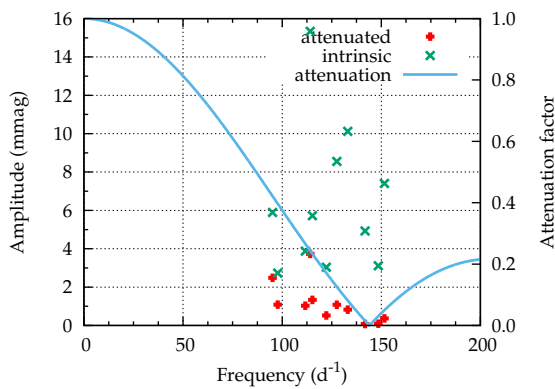}
\caption{The amplitudes and frequencies of the simulation data. The intrinsic amplitudes are shown with the green crosses. Due to 10-min exposure, the observed amplitudes are reduced as shown with red asterisks. }
\label{fig:12}
\end{figure}

\subsection{Numerical simulation with PLATO parameters}
\label{sec:7.1}

We demonstrate how the periodically-modulated sampling interval with these parameter values is useful to identify the true frequencies, by carrying out numerical simulations with artificial pulsation data similar to a rapidly oscillating Ap star. The star is assumed to be rotating with a period of 3.8\,d, and the pulsations are dipole axisymmetric high-order p modes, whose symmetry axis is oblique to the stellar rotation axis. 
As the star rotates, the aspect angle of the pulsation axis varies. Hence the apparent pulsation amplitudes are modulated with the rotation, and the amplitude spectrum is composed of triplets,  
\begin{eqnarray}
	\!\!\!\!\!\!\!\!\!\!
	A(t)\!\!\!\!\!\! &=& \!\!\!\!\!\!\sum_i A_i 
	\left[ \cos(2\uppi\nu_i t) 
	\right.
	\nonumber \\
	&+&\!\!\!\!\!\!
	\left.
	\alpha
	\left\{\cos(2\uppi (\nu_i\!+\!\nu_{\rm rot}) t) \!+\! \cos(2\uppi (\nu_i\!-\!\nu_{\rm rot}) t)\right\}
	\right]\!,
\label{eq:24}
\end{eqnarray}
where $A_i$ and $\nu_i$ denote the amplitude and the frequency of the $i$-th pulsation mode, respectively, and $\nu_{\rm rot}$ is the stellar rotation frequency. The coefficient $\alpha$ is determined by the the angle of the rotational axis to the line-of-sight and that of the rotation axis to the pulsation axis {\citep{kurtz1982}}.
The intrinsic amplitudes and frequencies of the simulation data are shown in Fig.\,\ref{fig:12} by green crosses. The value of $\alpha$ is $0.2$ in the simulation.

\begin{figure*} 
\centering
\includegraphics[width=0.8\columnwidth]{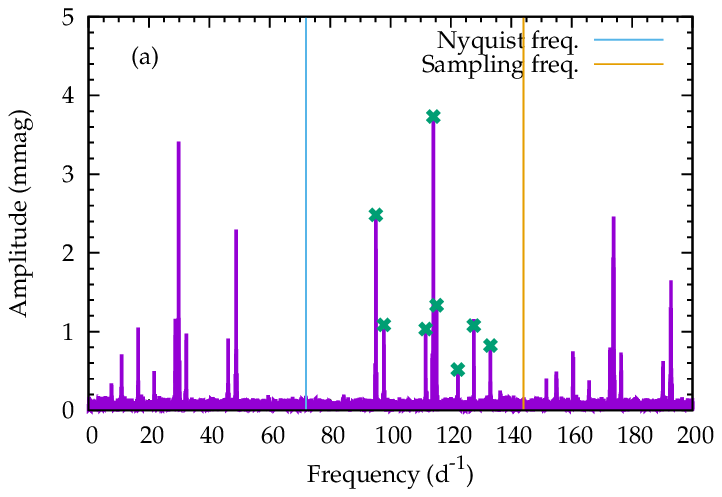}
\includegraphics[width=0.8\columnwidth]{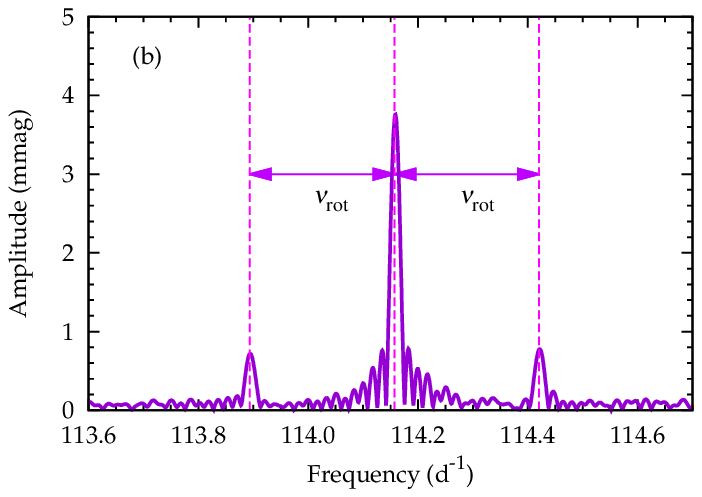}
\includegraphics[width=0.8\columnwidth]{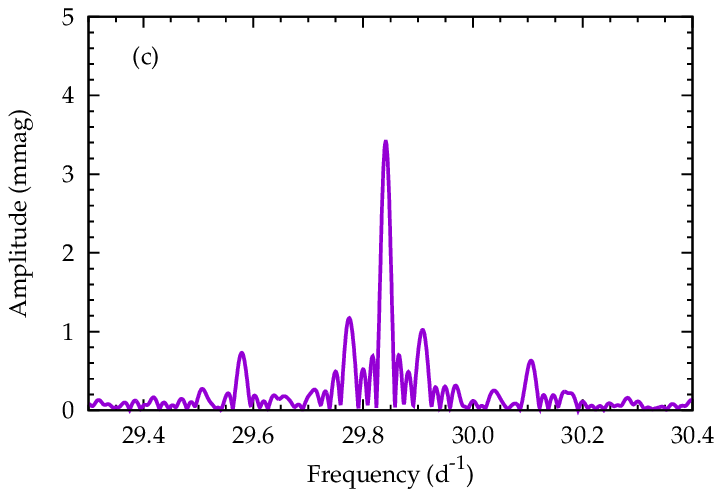}
\includegraphics[width=0.8\columnwidth]{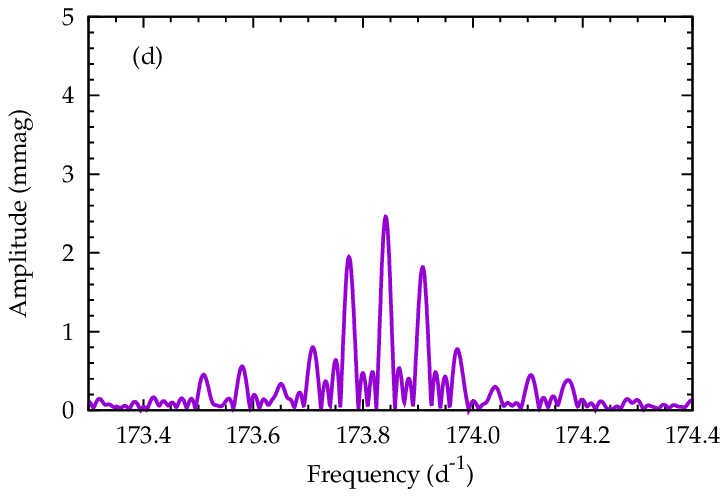}
\caption{{(a)} The amplitude diagram obtained with the Fourier transform of the simulation data with 10-min exposure in the frequency range of $0$\,-\,$200\,{\rm d}^{-1}$.  
{(b)} A zoom-in plot near the frequency range of the highest peak. 
{(c)} A zoom-in plot around the frequency range of the first order alias of the highest peak. 
{(d)} A zoom-in plot around the frequency range of the second order alias of the highest peak.}
\label{fig:13}
\end{figure*}

\subsubsection{Long exposure case}
\label{sec:7.1.1}

We first assume an exposure time of 10\,min with 10-min sampling. The amplitude reduction due to this long exposure is shown by the sinc function illustrated with a light blue curve in Fig.\,\ref{fig:12}, and the attenuated amplitude distribution is shown by red crosses.   
Figure\,\ref{fig:13}a shows the amplitude diagram obtained with the Fourier transform of the simulation data with 0.2-mmag level white noise, with 10-min exposure in the frequency range of $0$\,-\,$200\,{\rm d}^{-1}$. 
The amplitudes are attenuated significantly near the sampling frequency, and consequently oscillations near the sampling frequency are not detected.  
{Panel (b)} shows a zoom-in plot near the frequency range of the highest peak. The triplet with a separation of $\nu_{\rm rot} = 1/3.8\,{\rm d}^{-1}$ is clearly seen.
Figure\,\ref{fig:13}c shows a zoom-in plot around frequency range of the second highest peak, corresponding to the difference between the sampling frequency and the highest peak frequency. The central peak of the triplet caused by stellar rotation is clearly split as a new triplet with a separation of $\nu_{\rm mod}=1/15\,{\rm d}^{-1}$. Each of the side peaks of the triplet caused by stellar rotation are also split as a new triplet. Hence the peak around 29.8\,d$^{-1}$ is easily identified as the first alias of the true mode around 114.2\,d$^{-1}$. 
Figure\,\ref{fig:13}d shows a zoom-in plot around 173.8\,d$^{-1}$ indicating a quintuplet with a separation of $\nu_{\rm mod}$. The sidelobes separated from the central peak by 1/3.8\,d$^{-1}$ are seen as triplets rather than quintuplets, since the amplitudes of the outer most components of the quintuplets are too low. Despite that, a quintuplet with a separation of $\nu_{\rm mod}$ around the central peak is clearly seen, hence the peak around 173.8\,d$^{-1}$ is identified as a second alias of the true mode around 114.2\,d$^{-1}$.  

\begin{figure*}[t!] 
\centering
\includegraphics[width=0.8\columnwidth]{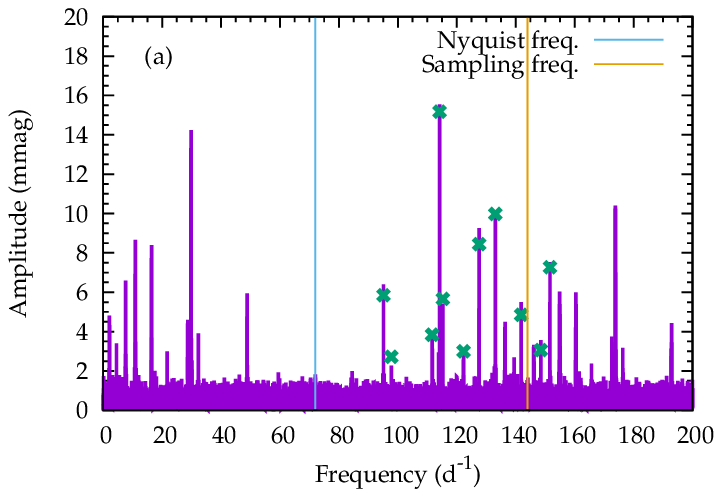}
\includegraphics[width=0.8\columnwidth]{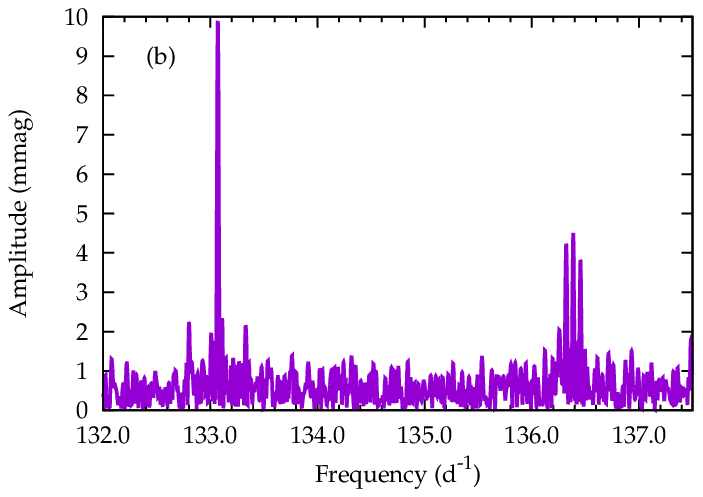}
\includegraphics[width=0.8\columnwidth]{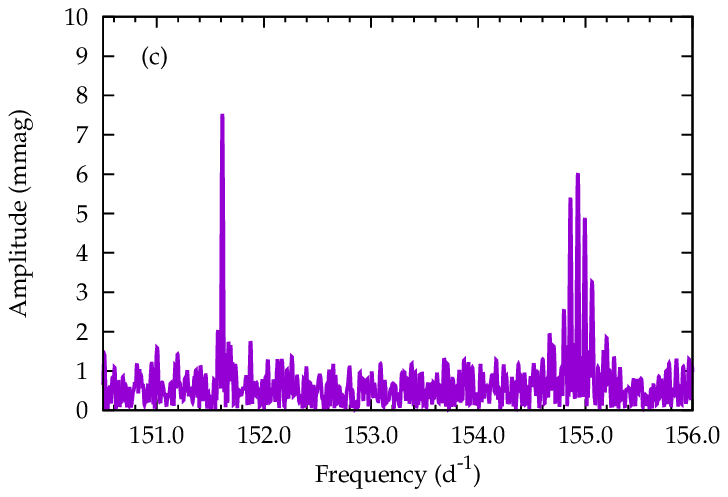}
\includegraphics[width=0.8\columnwidth]{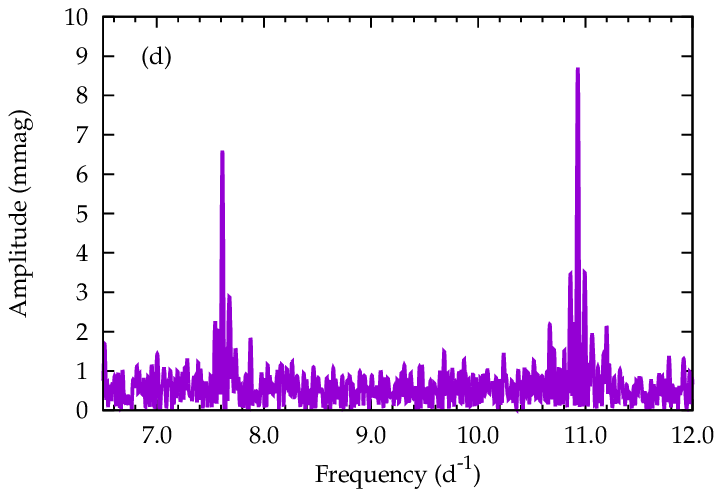}
\caption{{(a)} The amplitude diagram obtained with the Fourier transform of the simulation data with 50-sec exposure in the frequency range of $0$\,-\,$200\,{\rm d}^{-1}$.  
{(b)} A zoom-in plot near the frequency range of $132$\,-\,$137.5\,{\rm d}^{-1}$. The peak around $133\,{\rm d}^{-1}$ is a singlet, indicating that it is a true mode, while the peak around $136.5\,{\rm d}^{-1}$ is a quintuplet indicating that it is a second order alias of another mode. 
{(c)} A zoom-in plot around the frequency range of $150.5$\,-\,$156\,{\rm d}^{-1}$. The peak around $151.5\,{\rm d}^{-1}$ is a true frequency, as evidenced by being a singlet, while the peak around $155\,{\rm d}^{-1}$ is a quintuplet of a second-order alias of the true mode at $133\,{\rm d}^{-1}$. 
{(d)} A zoom-in plot around the frequency range of $6.5$\,-\,$12\,{\rm d}^{-1}$. Both of the peaks around $7.5\,{\rm d}^{-1}$ and $11\,{\rm d}^{-1}$ are triplets, indicating that they are first-order aliases. }
\label{fig:14}
\end{figure*}

\subsubsection{Short exposure case}
\label{sec:7.1.2}

As the second numerical test, we assume that the 600\,s sampling is divided into two exposure runs, the short exposure of 50\,s and the remaining 550\,s for a longer exposure so that the short exposure measurement with little attenuation is done every 600\,s. The modulation parameters are the same as in the previous case; $\nu_{\rm mod}=1/15\,{\rm d}^{-1}$ and $\tau_{\rm mod}=60\,{\rm s}$. 

The Fourier spectrum of the generated time series is shown in Fig.\,\ref{fig:14}a. The intrinsic amplitudes of the modes, marked with the green crosses, are well reproduced, as the attenuation is small.
{Panel (b)} shows a zoom-in plot around 135\,d$^{-1}$, a little bit lower than the sampling frequency $\nu_{\rm s}=144\,{\rm d}^{-1}$. A triplet composed of sharp peaks with a separation of $\nu_{\rm rot}=1/3.8\,{\rm d}^{-1}$ is seen around 133.0\,d$^{-1}$. The triplet is then identified as a true one. On the other hand, the peaks around 136.5\,d$^{-1}$ form a quintuplet, whose separation is $0.067\,{\rm d}^{-1} = \nu_{\rm mod}$, hence they are a second alias of another mode. 
{Panel (c)} shows a zoom-in plot around 153\,d$^{-1}$, which is the frequency range of mirror symmetry of {panel (b)} with respect to the sampling frequency $\nu_{\rm s}$.
The peak around 151.5\,d$^{-1}$ is located at the mirror symmetric frequency of the quintuplet shown in {panel (b)} and it is a singlet, which vaguely seems to be a triplet with a separation of $1/3.8\,{\rm d}^{-1}$, indicating it as a true mode. Indeed the frequency of the central peak of the quintuplet in {panel (b)} is equal to the difference between the frequency twice the sampling frequency and the frequency of the singlet in {panel (c)}, indicating it as a second alias of the true mode.  
The peaks around 155.0\,d$^{-1}$ form a quintuplet with a separation of $\nu_{\rm mod}$, and a pair of peaks separated from the central peak of the quintuplet by  $1/3.8\,{\rm d}^{-1}$ are also seen.
These features indicate that this quintuplet is a second alias of the true mode appeared in {panel (b)} at 133.0\,d$^{-1}$. 

Once we identify the true peaks around 133\,d$^{-1}$ and 151.5\,d$^{-1}$, we expect that their first aliases appear as triplets, respectively, at the frequencies corresponding to the difference between them and the sampling frequency. Figure\,\ref{fig:14}d confirms that this is the case. Both peaks around 7.5\,d$^{-1}$ and 11.0\,d$^{-1}$ are triplets with a separation of $\nu_{\rm s}$ and they have a pair of peaks separated from the central peak by $1/3.8\,{\rm d}^{-1}$.

\subsection{Summary of recommendation for PLATO}
\label{sec:7.2}

From these numerical simulations, in order to maximise capability of PLATO's 600\,s cadence photometry for asteroseismology, we propose (i) to introduce a periodic modulation to the sampling rate with parameters such as $\nu_{\rm mod}=1/15\,{\rm d}^{-1}$ and $\tau_{\rm mod}=60\,{\rm s}$ and (ii) to split the cadence into a short exposure of 50\,s and 550\,s.

\section{Conclusion}
\label{Sec:8}

Due to limits on the telemetry of space missions, the sampling rate of the photometry is rarely as high as one would wish. The resulting Nyquist aliases of oscillations quite often complicate the subsequent data analysis.
  
We have shown analytically and with numerical simulations, that introduction of an appropriate modulation to the sampling rate makes it possible to identify correctly the true oscillation frequencies even if they are higher than the Nyquist frequency. 
The appropriate value for the frequency of the modulation, $\nu_{\rm mod}$, is  $\gtrsim 4$ times as large as either the reciprocal of the time span of observations or the damping rate of stochastically excited acoustic modes of stars.  
For the amplitude, $\tau_{\rm mod}$, we suggest roughly one tenth of the averaged sampling interval. 

In order to reproduce the intrinsic amplitude spectrum, short exposures are favourable. 
Even though the noise level in short exposures is greater than for full-cadence exposures, 
the former case has less amplitude smear and measures a higher pulsation amplitude, compensating for the higher noise.
The remaining time in the long cadence can still be used for long exposure for planet hunting and asteroseismology for long period pulsators and the two sub-exposures can be recombined after download if desired.

In summary,
\begin{enumerate}
\item
by modulating the sampling rate, we can overcome the problem of the Nyquist aliases and enhance the already-revolutionary photometry from space, 
\item
by dividing the cadence into two unequal parts and by analysing separately the photometry data taken with the shorter runs, we can reproduce the intrinsic amplitude spectrum, which is informative for asteroseismology,

and
 
\item
these are highly effective to maximise the capability of PLATO's photometry for asteroseismology, without seriously interfering with search for exoplanets, another very important scientific mission of PLATO. 
\end{enumerate}

\section*{Acknowledgments}
The authors thank the organizing committees of the conference, led by Sylvie and Gerard Vauclair, for their excellent organization and good atmosphere of this fruitful meeting. This work was partly supported by the JSPS Grant-in-Aid for Scientific Research (16K05288). 

\bibliographystyle{phostproc}
\bibliography{sna}

\renewcommand{\theequation}{A\arabic{equation}}
\setcounter{equation}{0}
\section*{Appendix 1: Fourier transform of discretely sampled data}

We consider a series of discretely sampled data of the stellar brightness  $f_N(t)$: 
\begin{equation}
	f_N(t)=\sum_{n=1}^N f(t) \delta(t-t_n),
\label{eq:A1}
\end{equation}
where $f(t)$ represents a continuous variation of the stellar brightness in time and 
$\delta(t-t_n)$ is Dirac's delta function.
The Fourier transform of $f_N(t)$ is defined by
\begin{equation}
	F_N(\nu) := \int_{t_1}^{t_N} f_N(t) {\rm e}^{-2\uppi {\rm i}\nu t}\, {\rm d}t.
\label{eq:A2}
\end{equation}
By substituting (\ref{eq:A1}) into (\ref{eq:A2}) after using the following expression of the delta function,
\begin{equation}
	\delta(t-t_n) = \int_{-\infty}^{\infty} {\rm e}^{2\uppi {\rm i} \nu'(t-t_n)}\, {\rm d}\nu' ,
\label{eq:A3}
\end{equation} 
and changing the order of integrals, we obtain $F_N(\nu)$ in the form of a convolution of the Fourier transform of $f(t)$ and the sampling window spectrum $W_N(\nu)$: 
\begin{eqnarray}
	{{F_N(\nu)}\over{N}} 
	\!\!\!\!\!&=&\!\!\!\!\!
	\int_{t_1}^{t_N} \!\!\!f(t)\!
	\left[ {{1}\over{N}}\sum_{n=1}^N 
	\int_{-\infty}^{\infty} \!\!{\rm e}^{2\uppi {\rm i} \nu' (t-t_n)}\, {\rm d}\nu'\right]
	{\rm e}^{-2\uppi {\rm i}\nu t}\, {\rm d}t
	\nonumber\\
	\!\!\!\!\!&=&\!\!\!\!\!
	\int_{-\infty}^{\infty}\! 
	\left[ \int_{t_1}^{t_N} \!\!\!f(t) {\rm e}^{-2\uppi {\rm i}(\nu-\nu')t} {\rm d}t \right]
	\!{{1}\over{N}}\! \sum_{n=1}^N {\rm e}^{-2\uppi{\rm i}\nu' t_n} {\rm d}\nu'
	\nonumber\\
	\!\!\!\!&=&\!\!\!\!
	\int_{-\infty}^{\infty} F(\nu-\nu') W_N(\nu') \,{\rm d}\nu',
\label{eq:A4}
\end{eqnarray}
where
\begin{equation}
	F(\nu) := \int_{t_1}^{t_N} f(t) {\rm e}^{-2\uppi {\rm i}\nu t}\, {\rm d}t
\label{eq:A5}
\end{equation}
is the Fourier transform of the continuous function $f(t)$ and 
\begin{equation}
	W_N(\nu) := {{1}\over{N}}\sum_{n=1}^N {\rm e}^{-2\uppi{\rm i}\nu t_n}.
\label{eq:A6}
\end{equation}

\smallskip
\section*{Appendix 2: Window spectrum of periodically modulated sampling}
Subsituting the data sampling time 
\begin{equation}
	t_n = t_0+n{ {1}\over{\nu_{\rm s}} } 
	+ \tau_{\rm mod} \sin \left( 2\uppi n {{\nu_{\rm mod}}\over{\nu_{\rm s}}} \right)
\label{eq:A7}
\end{equation}
into equation (\ref{eq:A6}) and making the square, we get, with the help of the generating function of the Bessel function
\begin{equation}
	\exp(-{\rm i}z\sin \theta) 
	= \!\! \sum_{k=-\infty}^{\infty} J_k(z)\exp(-{{\rm i}k\theta}),
\label{eq:A8}
\end{equation}
\begin{eqnarray}
	\left|W_N(\nu)\right|^2
	\!\!\!\!\!\!&=&\!\!\!\!\!
	{{1}\over{N^2}}
	\left| \sum_{n=1}^N \left[
	\exp \left\{ -{\rm i} \left( 2\uppi n {{\nu}\over{\nu_{\rm s}}} \right) \right\} \right.\right.
	\nonumber\\
	\!\!\!\!\!\!&&\!\!\!\!\!\!
	\times
	\left.\left.
	\left[ \sum_{k=-\infty}^\infty \!\!J_k(\xi) 
	\exp \left\{ -{{\rm i} k \left(2\uppi n {{\nu_{\rm mod}}\over{\nu_{\rm s}}} \right) } \right\}
	\right]\right]
	\right|^2
	\nonumber\\
	\!\!\!\!\!&=&\!\!\!\!\!
	{{1}\over{N^2}}\!
	\left| \! \sum_{k=-\infty}^\infty \!\!\!J_k(\xi) \!
	{{1\!-\!\exp \! \left\{ \! -{\rm i}
	2N\uppi \! \left( \! {{\nu}\over{\nu_{\rm s}}}\!+\!k{{\nu_{\rm mod}}\over{\nu_{\rm s}}}\! \right)\!\!\right\}}
	\over{1\!-\!\exp \! \left\{ \! -{\rm i}
	2\uppi \! \left( \! {{\nu}\over{\nu_{\rm s}}}\!+\!k{{\nu_{\rm mod}}\over{\nu_{\rm s}}}\! \right)\!\!\right\}
	}}
	\right|^2
	\nonumber\\
	\!\!\!\!\!&=&\!\!\!\!\!
	{{1}\over{N^2}}
	\left| \sum_{k=-\infty}^\infty \!\!
	{\rm e}^{-{\rm i}(N-1)\uppi k{{\nu_{\rm mod}}\over{\nu_{\rm s}}}} \right.
	\nonumber\\
	\!\!\!\!\!&&\!\!\!\!\!\!
	\times 	
	\left.
	J_k(\xi) 
	{{\sin \left[ N\uppi \left\{ \left( {\nu} + k {\nu_{\rm mod}} \right) / {\nu_{\rm s}} \right\}  \right] }
	\over
	{\sin \left[ \uppi \left\{ \left( {\nu} + k {\nu_{\rm mod}} \right) / {\nu_{\rm s}} \right\}  \right] }}
	\right|^2,
\label{eq:A9}
\end{eqnarray}
where
\begin{equation}
	\xi := 2\uppi\nu\tau_{\rm mod}.
\label{eq:A10}
\end{equation}
Note that the order of summations was changed in the middle of equation (\ref{eq:A9}).
The Bessel coefficients of different orders are orthogonal each other. 
As a consequence, the cross terms of the Bessel coefficients with different orders are much smaller than those of the squared terms. Hence,
\begin{equation}
	|W_N(\nu)|\simeq
	{{1}\over{N}} \sum_{k=-\infty}^\infty 
	\left| J_k(\xi) 
	{{\sin \left[ N\uppi \left\{ \left( {\nu} + k {\nu_{\rm mod}} \right) / {\nu_{\rm s}} \right\}  \right] }
	\over
	{\sin \left[ \uppi \left\{ \left( {\nu} + k {\nu_{\rm mod}} \right) / {\nu_{\rm s}} \right\}  \right] }}
\right|.
\label{eq:A11}
\end{equation}

\end{document}